# Theoretical investigation on armchair graphene nanoribbons with oxygen-terminated edges

Hongyu Ge, Guo Wang* and Yi Liao

*Department of Chemistry, Capital Normal University, Beijing 100048, China*

*Corresponding author: Tel.: +861068902974; E-mail address: wangguo@mail.cnu.edu.cn

ABSTRACT

Armchair graphene nanoribbons with different proportions of edge oxygen atoms are investigated by using crystal orbital method based on density functional theory. All the nanoribbons are energetically favorable, although buckled edges are present. Isolated edge oxygen atoms cause semiconductor-metal transition via introducing edge states, while adjacent edge oxygen atoms not. For the graphene nanoribbons with all oxygen atoms on the edges, both band gap and carrier mobility alternate with respect to the ribbon width. The carrier mobilities are as 18%-65% large as those of the graphene nanoribbons with hydrogen-terminated edges. These values are as large as $10^3$ cm$^2$·V$^{-1}$·s$^{-1}$, which are still quite high for electronic devices. Crystal orbital analysis gives pictorial explanations to the phenomenon.

## 1. Introduction

Graphene has attracted extensive attention for its outstanding mechanical and electronic properties [1]. Graphene could be a good candidate for next-generation electronic devices because of its extremely high carrier mobility [2,3]. Finite width of graphene nanoribbons (AGNR) [4,5] can open up zero band gap of graphene [6], which would produce high on/off ratio and then benefit real applications.

Bottom-up approaches can synthesize atomically precise graphene nanoribbons with hydrogen-terminated edges [7]. This gives a possibility of probing intrinsic properties of graphene nanoribbons. However, only 7-AGNR is synthesized. On the other hand, oxygen plasma etching in top-down lithographic method can also produce graphene nanoribbons [5]. Oxygen atoms could likely to passivate their active edges. Moreover, ketone structures can also present at edge of graphene oxide [8], which is prepared by treating graphite with strong oxidizing acids. Compared with graphene, significant degradation of carrier mobilities occurs for graphene nanoribbons [9-12], which is explained by edge roughness, defects and phonons [10,13]. However, the influence coming from edge oxygen atoms has not been investigated so far. It is important to make clear that whether high carrier mobilities maintain for the graphene nanoribbons with edge oxygen atoms. It is also interesting to investigate the influence from the edge oxygen atoms on the electronic properties of graphene nanoribbons.

Theoretical investigations indicated that zigzag graphene nanoribbons with oxygen-terminated edges have almost zero or zero band gaps [14,15], which should have difficulty in achieving high on/off ratio in operation of electronic devices. In this work, armchair graphene nanoribbons with different proportions of oxygen atoms presented at the edges are investigated by using crystal orbital method based on density functional theory. It is shown that the edge oxygen atoms could introduce semiconductor-metal transition for the graphene nanoribbons, and the carrier mobilities are analyzed under deformation potential theory.

## 2. Models and computational details

Primitive cell and supercell of 9-AGNR are used to investigate the influence from edge oxygen atoms with different proportions. We use 1cell-1 to denote the primitive cell in Fig. 1(a) with $H_1$ replaced by an oxygen atom, while 2cell-12 indicates the supercell in Fig. 1(b) with $H_1$ and $H_2$ replaced by two oxygen atoms. The structures are fully optimized by using crystal orbital method with CRYSTAL06 program [16]. Hybrid density functional B3LYP(VWN5) [17,18] and Bloch functions constructed with 6-21G(*d*) basis set are used throughout the calculations. Default values of convergence criteria in the program are used. A Monkhorst-Pack sampling with 41 $k$-points in the first Brillouin zone is adopted in the self-consistent iteration procedure and is sufficient to obtain converged energy and derived electronic properties. A uniform $k$-points sampling with 800 points is used in the non-iterative band structure calculation in order to facilitate the fitting of the parameters.

Under deformation potential theory, carriers are assumed to be scattered mostly by acoustic phonons [19]. For these one-dimensional oxygen-terminated AGNR, carrier mobilities are obtained by

$$\mu = \frac{e\hbar^2 C}{(2\pi k_B T)^{1/2} |m^*|^{3/2} E_1^2}, \quad (1)$$

where $C = a_0 \partial^2 E / \partial a^2 |_{a=a_0}$, $a_0$ is lattice parameter of one-dimensional structure, $E_1 = a_0 \delta \varepsilon / \delta a$, $\varepsilon$ is energy for top of valence band or bottom of conduction band, and $m^* = \hbar^2 [\partial^2 E / \partial k^2]^{-1}$. Stretching modulus $C$ and deformation potential constant $E_1$ in the process of scattering by longitudinal acoustic phonon along one-dimensional direction are obtained under deformed geometries ($a$=0.99, 0.995, 1.005 and 1.01 $a_0$) [20], carrier effective mass $m^*$ is obtained by fitting frontier bands. It is noted that carriers without a very sharp density of state near edge of frontier bands could contribute to the real conduction process in an energy range wider than the thermal energy $k_B T$. In this work, $10 k_B T$ [21] is used to obtain the carrier effective mass and then the carrier mobility. The deformation potential approach has been successfully applied to one-dimensional graphene nanoribbons [22-24] and carbon nanotubes [25,26]. The carrier mobility of graphene calculated under deformation potential theory [24] agrees well with the experimental result [27].

## 3. Results and discussion

### 3.1 Structures and stabilities

All the optimized structures have closed shell configurations. This is further verified with spin-polarized calculations by using VASP program [28]. Planar geometries maintain in the middle of the nanoribbon, while buckled edges are found especially when two oxygen atoms are linked to adjacent carbon atoms (1cell-13, 1cell-123, 1cell-1234, 2cell-13, 2cell-123, 2cell-136). In the supposed planar structures (all atoms are in a plane), the distances of the adjacent edge oxygen atoms are as short as 1.4-1.5 angstrom, which are too near for non-bonded oxygen atoms. It is also indicated that the planar structures with all oxygen atoms on the edges are not stable from VASP calculations [29]. The stress can be released when oxygen atoms deviate to different directions. The buckled structure of 1cell-1234 is shown in Fig. 1(c). The two oxygen atoms at the edge deviate from the original plane of the graphene nanoribbons. For the edge carbon atoms, the deviation from the plane is about $10°$.

The formation of graphene nanoribbons with hydrogen-terminated edges are energetically favorable compared with parent graphene and hydrogen [30]. For graphene nanoribbons with dangling bonds at edges, edge oxygen atoms should passivate active bonds and always make them stable. The following reaction, for example,

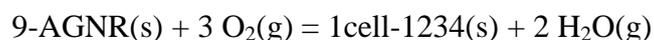
$$9\text{-AGNR(s)} + 3\ O_2(g) = 1\text{cell-1234(s)} + 2\ H_2O(g)$$

is used to compare the stabilities of the AGNRs and the oxygen-terminated AGNRs. In the above reaction, the oxygen-terminated structure has four oxygen atoms at the two edges, the $\Delta E$ (defined as energy change per edge oxygen atom) is -1.59 eV, indicating that the edge-oxidization is energetically favorable. The $\Delta E$ for other oxygen-terminated structures are calculated in similar ways. From Table 1, it can be seen that the edge-oxidizations are all favorable. For the primitive cell, the formation of 1cell-1234 is the most exothermic. For the primitive cells with two edge oxygen atoms (edge oxygen proportion is 1/2), the structure 1cell-13 is the most stable. The $\Delta E$ is -1.56 eV, while it is -0.94 or -0.98 eV for 1cell-12 or 1cell-14. It is unexpected that the structure 1cell-13 with deformed edges is more stable. This will be discussed later. Some structures in Table 1 constructed with supercells are also calculated to investigate more edge oxygen proportions. It is similar that the structure 2cell-13 with adjacent oxygen atoms and deformed edge is more stable than the structure 2cell-12 with isolated oxygen atoms.

Since it is energetically favorable for each replacement of a hydrogen atom by an oxygen atom, the structure with all oxygen atoms on the edges is the most stable. In order to investigate the variation of properties with the nanoribbon width $N$, AGNR with all edge hydrogen atoms replaced by oxygen atoms (denoted as $N$-AGNR-O, $N$=10-18) are also calculated. The optimized $N$-AGNR-O also have buckled geometries at the two edges. The deviation of the two carbon atoms at each edge is about 10° for all the AGNR-O, while planar geometries maintains in the middle. This is similar to the situation for 9-AGNR-O, i.e. 1cell-1234. The range of $\Delta E$ for these nanoribbons is from -1.61 to -1.58 eV. The reason of the narrow range is that the oxidization and the buckled structures are localized at the edges, and the situation does not change with the width.

**3.2 Electronic properties of 9-AGNR with oxygen-terminated edges**

The oxygen-terminated AGNRs fall into three categories according to their band structures. For the primitive cell, the first category incorporates 1cell-13, 1cell-1234 and all the AGNR-O. These structures have even number of oxygen atoms, and meanwhile the oxygen atoms are linked to adjacent edge carbon atoms. As an example, the band structures of 1cell-1234 are shown in Fig. 2(a). These structures are all semiconductors, and direct band gaps exist at the center of the first Brillouin zone. The second category includes 1cell-1, 1cell123. Although 1cell123 has stable two adjacent oxygen atoms, there is still another isolated oxygen atom. These structures with odd number of oxygen atoms have metallic band structures. As an example, the band structures of 1cell-1 are shown in Fig. 2(b), in which there is a half-filled band. When there is odd number of edge oxygen atom(s), the number of electrons in a cell should be also odd. It is natural that the structures have half-filled bands. The last category incorporates 1cell-12 and 1cell-14. The band structures of 1cell-12 are shown in Fig. 2(c). Although these two structures have even number of oxygen atoms, the band structures have still metallic character with two partial filled bands.

In order to explain the three categories, frontier crystal orbitals are constructed and analyzed. The unpaired electron should exist near the oxygen atom in 1cell-1. The extension of this edge state is localized at and near the oxygen atom shown in Fig. 3(a) and 3(b). The combination of oxygen p orbital and edge carbon p orbitals has high energy, which composes the frontier crystal orbitals at the Fermi level. The narrow band width is 0.83 eV due to the edge nature, so the density of states (DOS) is high (1.32 states per eV listed in Table 1). For 1cell-123, the frontier crystal orbitals in Fig. 3(c) and 3(d) are also localized edge states generated by the isolated oxygen atom.

The highest occupied crystal orbital (HOCO) and the lowest unoccupied crystal orbital (LUCO) of 1cell-13 in Fig. 3(e) and 3(f) are delocalized orbitals involving $\pi$ states. The two unpaired electrons generated by two adjacent oxygen atoms should make a pair with each other. The energy of the occupied edge state should decease and the corresponding band does not appear in the frontier states.

The frontier crystal orbitals of 1cell-12 at $\alpha$ or $\beta$ point (labeled in Fig. 2(c)) are shown in Fig. 3(g) and 3(h) or 3(i) and 3(j), respectively. The image part at $\alpha$ point in Fig. 3(h) and the real part at $\beta$ point in Fig. 3(i) have edge character. The real part at $\alpha$ point in Fig. 3(g) and the image part at $\beta$ point in Fig. 3(j) are combinations of isolated p atomic orbitals. Since the two oxygen atoms are not linked to adjacent edge carbon atoms, there should exist two edge states. The linear combination of the two states as well as some isolated p atomic orbitals generates two sets of frontier crystal orbitals. From the Figures, it can be seen that the energies at these two points should be similar, and there exist two partial filled frontier bands for 1cell-12.

The occupied edge localized state has higher energy than delocalized state has. It is rational that 1cell-12 or 1cell-14 is less stable than 1cell-13. Although 1cell-13 has buckled edges, the pair of high energy electrons should release more energy and makes structures with adjacent edge oxygen atoms more stable. It is not unexpected that the structure 1cell-13 with deformed edges is more stable from this point of view.

For the supercells, the situation is similar. The structure 2cell-13 with two adjacent oxygen atoms has semiconductor band structures. The structures 2cell-1, 2cell-123 and 2cell-126 with odd number of oxygen atoms have half-filled metallic bands. The structure 2cell-12 has two partial filled bands. It is also similar that 2cell-12 with isolated oxygen atoms is less stable than 2cell-13 with adjacent oxygen atoms.

Isolated edge oxygen atoms could introduce semiconductor-metal transition in AGNRs. The DOS listed in Table 1 are in the range of 1.10-1.43 state per eV per primitive cell for 1cell-1, 1cell123, 2cell-1, 2cell-123 and 2cell-136 with a half-filled band. The DOS increase to 2.15-2.70 state per eV per primitive cell for 1cell-12, 1cell-14 and 2cell-12 with two partial filled bands. Although the DOS at Fermi level are an order larger than that of a single-walled carbon nanotube [31], they are still an order smaller than that of doped $C_{60}$ [32]. The electron-phonon superconducting temperature should be not high compared with doped $C_{60}$. They are not suitable for high-temperature superconducting materials according to electron-phonon coupling theory. Furthermore, the metallic structures are less stable than the semiconducting structures. Since the semiconducting structures with all oxygen atoms on the edges are the most stable, electronic properties of these structures are discussed below.

### 3.3 Electronic properties of *N*-AGNR-O

All the band gaps exist at the center of the first Brillouin zone for the AGNR-O, indicating that they are all semiconductors. The band gaps shown in Fig. 4(a) oscillate with the width *N*. The band gaps of 10-AGNR-O, 13-AGNR-O and 16-AGNR-O are all smaller than 0.3 eV, while for others the band gaps are in the range of 0.8-1.4 eV. Like hydrogen-passivated AGNR, AGNR-O can fall in to three groups: $N=3q$, $N=3q+1$ and $N=3q+2$, where $q$ is a positive integer. The band gap sequence of AGNR-O is $gap_{3q} > gap_{3q+2} > gap_{3q+1}$. This is in agreement with the results calculated by VASP program [29] and is different from the sequence $gap_{3q+1} > gap_{3q} > gap_{3q+2}$ of AGNR [4,23]. Normally, the oxidation could open up the band gap. For example, from 14-AGNR to 14-AGNR-O, the band gap increases from 0.20 eV to 1.04 eV. However, this is not always true for the graphene nanoribbons. From 13-AGNR to 13-AGNR-O, the band gap decreases from 1.22 eV to 0.09 eV. This is due to the different band gap sequences of AGNR-O and AGNR. The reason will be discussed later. Nevertheless, the band gap decreases with the ribbon width, and should reach zero for two-dimensional graphene.

The stretching modulus of 9-AGNR-O is 270 eVÅ$^{-1}$, and this value increases gradually to 544 eVÅ$^{-1}$ for 18-AGNR-O. The stretching moduli are as 83%-95% magnitude as those the corresponding AGNR [23] have (calculated with the same density functional and basis set), indicating slight decrement of the mechanical performance due to the buckled edges. Nevertheless, the elastic constant should be also in the order of 1 TPa, which the AGNR have.

The hole and electron effective masses within 10 $k_BT$ from the frontier band edges are in the range of 0.09-0.22 and 0.10-0.35 $m_0$ for $N$-AGNR-O ($N$=10-18), where $m_0$ is the mass of a free electron. For 9-AGNR-O, these values are 0.34 and 0.98 $m_0$, which are much larger due to the narrow $\pi$ conjugated system. The carriers in AGNR-O are heavier than those in AGNR (0.07-0.21 and 0.07-0.22 $m_0$ for hole and electron) [23], indicating that the edge-oxidation would decrease the carrier mobilities by enhancing the carrier masses.

Under deformed geometry along one-dimensional direction, deformation potential constant can be calculated from band edge shift with respect to lattice deformation proportion [20]. The deformation potential constant of the AGNR-O is unbalanced. For example, the valence or conduction band deformation potential constant ($E_{1v}$ or $E_{1c}$) of 9-AGNR-O is 9.3 or 6.3 eV, respectively. As shown in Fig. 4(b), $E_{1v}$ is larger than $E_{1c}$ when $N$=3$q$+2, while $E_{1c}$ is larger than $E_{1v}$ when $N$=3$q$ and 3$q$+1, with only exception of the narrow 10-AGNR-O. The "large" deformation potential constants are in the range of 9.3-12.0 eV, while the "small" ones are in the range of 5.0-7.1 eV. This will significantly affect the carrier mobility, because the carrier mobility depends on powers of the deformation potential constant in equation (1).

The reason why one type of deformation potential constant is larger than the other can be explained by crystal orbitals. Since deformation potential constant is related to the band edge shift, frontier crystal orbitals at the band edges should give pictorial descriptions. In Fig. 3(k) and 3(l), the HOCO and the LUCO of 9-AGNR-O are localized and delocalized with respect to the one-dimensional deformation direction, respectively. Upon deformation, the extended delocalized orbital along the one-dimensional conjugated line has smaller variation with the bond length change than the localized one. This results in smaller energy change of the delocalized orbital and then the smaller deformation potential constant. For other AGNR-O except 10-AGNR-O, delocalized orbitals exist at valence (conduction) band edge when $N=3q$ or $3q+1$ ($3q+2$). From two-dimensional graphene to one-dimensional AGNR-O, quantum confinement makes properties of frontier orbitals alternate, which results in unbalanced deformation potential constants.

It is noted that carbon atoms at the two edges of 9-AGNR-O shown in Fig. 3(k) and 3(l) contribute little to the frontier crystal orbitals due to the edge-oxidation and the buckled structure. The number of carbon atoms involved in the conjugation should not include the edge carbon atoms. Although p orbitals of edge oxygen atoms exist in the frontier crystal orbitals, they are isolated from the main conjugated orbitals in the middle. The electronic properties of $N$-AGNR-O should be similar to ($N$-2)-AGNR. For AGNR-O with $N=3q+2$, $E_{1v}$ is larger than $E_{1c}$; while for AGNR, this occurs when $N=3q$ [23]. Considering of the edge situation, the two rules should be the same. For band gap alternation discussed above, the difference between AGNR-O and AGNR should be due to the same reason.

For the exception 10-AGNR-O, the frontier orbitals shown in Fig. 3(m) and 3(n) are not distributed on all the middle carbon atoms. This narrow nanoribbon with a very small band gap has isolated conjugated structures and two carbon lines in the middle do not contribute to the frontier crystal orbitals. The total number of carbon atoms in the conjugated structures changes, so the deformation potential constants of 10-AGNR-O do not obey the above rule.

The carrier mobility shown in Fig. 4(c) oscillates with respect to the width $N$. The electron mobility $\mu_e$ is larger than the hole mobility $\mu_h$ when $N=3q+2$, because $E_{1c}$ is smaller than $E_{1v}$. For $N=3q$ or $3q+1$, the hole mobility is larger than the electron mobility. The only exception is 10-AGNR-O, due to the unusual isolated conjugated lines. The unbalanced deformation potential constants create the unbalanced carrier mobilities. The polarity [22] of the carrier mobilities depends on that of the deformation potential constants, and the different rules for the carrier mobility of AGNR-O and AGNR [23] can be considered to be the same, just like the situation for the deformation potential constant.

The carrier mobility generally increases with the width according to the larger stretching moduli and larger conjugated structures. The highest value 4096 $cm^2 \cdot V^{-1} \cdot s^{-1}$ occurs for the electron mobility of 17-AGNR-O, and this value is as 40% magnitude as that of 15-AGNR [23]. Weaker stretching modulus caused by edge-oxygen and edge-buckled structures, higher deformation potential constant, and slightly larger carrier effective mass are all responsible for the decrement. However, the decrement is not significant. The major carrier mobilities of $N$-AGNR-O are as 18%-65% large as those of ($N$-2)-AGNR. The major carrier mobilities are in the order of $10^3$ $cm^2 \cdot V^{-1} \cdot s^{-1}$, and are still quite high for nano-electronic devices.

## 4. Conclusions

Armchair graphene nanoribbons with oxygen-terminated edges are investigated based on density functional theory. Compared with hydrogen-terminated nanoribbons, all the nanoribbons with different proportions of edge oxygen atoms are stable, although buckled edges are present. The nanoribbons fall into three categories according to different band structures. The first has semiconductor band structures with adjacent edge oxygen atoms. The second has metallic band structures with a half-filled band, in which there is an isolated edge oxygen atom. The last has metallic band structures with two partial filled bands, where there are two isolated edge oxygen atoms. Edge states pictorial presented by crystal orbitals are responsible for the semiconductor-metal transition.

The band gap alternates with the width for the most stable armchair graphene nanoribbons with all oxygen atoms on the edges. The band gap sequence is $gap_{3q} > gap_{3q+2} > gap_{3q+1}$. The difference between the alternating rules for AGNR-O and AGNR are due to the buckled edges of AGNR-O, where no contribution is made to the frontier crystal orbitals. From AGNR to AGNR-O, the edge-oxidation increases or decreases the band gap and could change the polarity of the carrier mobility, according to different quantum confinements. The reason is that the edge carbon atoms with no contribution to the frontier crystal orbitals are excluded when counting the ribbon width. Unlike edge-roughness, edge oxygen atoms have limited negative impact on the carrier mobilities, and the highest mobilities are still in the order of $10^3$ $cm^2 \cdot V^{-1} \cdot s^{-1}$. The oxygen-terminated edges maintain high carrier mobility feature of armchair graphene nanoribbons as long as the edges are ordered.

**Table 1**

Energy change per edge oxygen atom (Δ$E$ in eV) from AGNR to edge-oxidized graphene nanoribbons, density of states per primitive cell at Fermi level (DOS in eV) of the edge-oxidized graphene nanoribbons, and edge oxygen proportions.

|            | Δ$E$  | DOS  | proportion |
|------------|-------|------|------------|
| 1cell-1    | -0.85 | 1.32 | 1/4        |
| 1cell-12   | -0.94 | 2.15 | 1/2        |
| 1cell-13   | -1.56 | 0    | 1/2        |
| 1cell-14   | -0.98 | 2.50 | 1/2        |
| 1cell123   | -1.35 | 1.38 | 3/4        |
| 1cell-1234 | -1.59 | 0    | 1          |
| 2cell-1    | -0.98 | 1.43 | 1/8        |
| 2cell-12   | -1.07 | 2.74 | 1/4        |
| 2cell-13   | -1.69 | 0    | 1/4        |
| 2cell-123  | -1.53 | 1.10 | 3/8        |
| 2cell-136  | -1.50 | 1.11 | 3/8        |

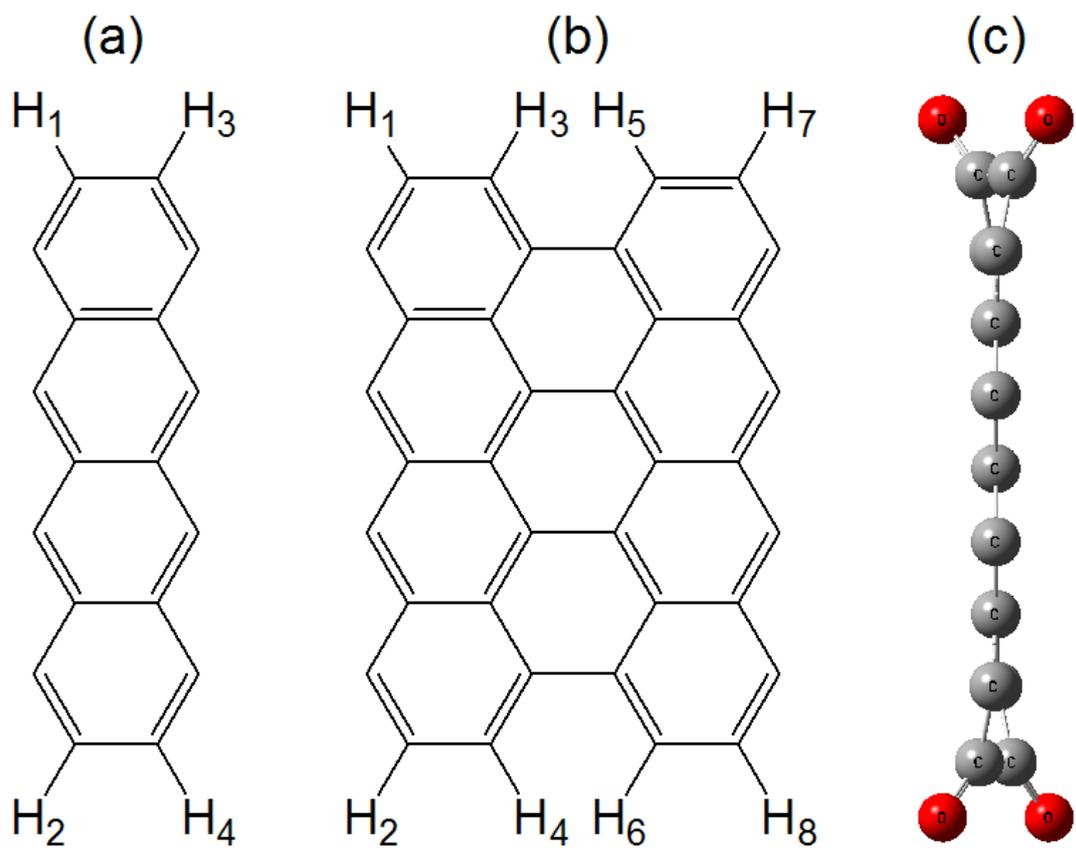

**Fig. 1.** Models of (a) primitive and (b) supercell of 9-AGNR, (c) optimized structure of 9-AGNR-O.

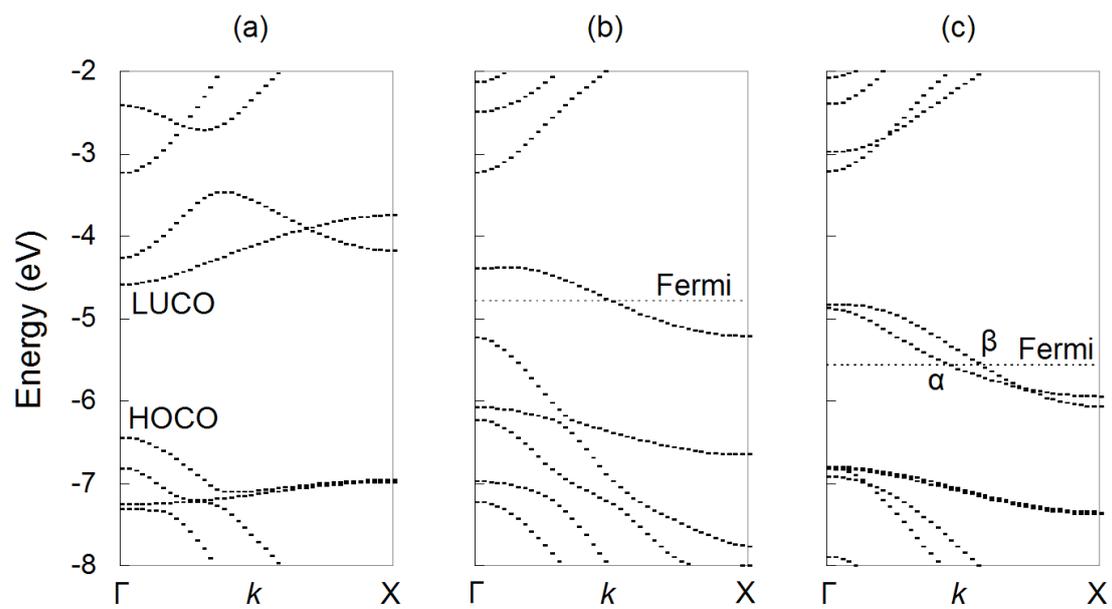

**Fig. 2.** Band structures of (a) 1cell-1234, (b) 1cell-1 and (c) 1cell-12.

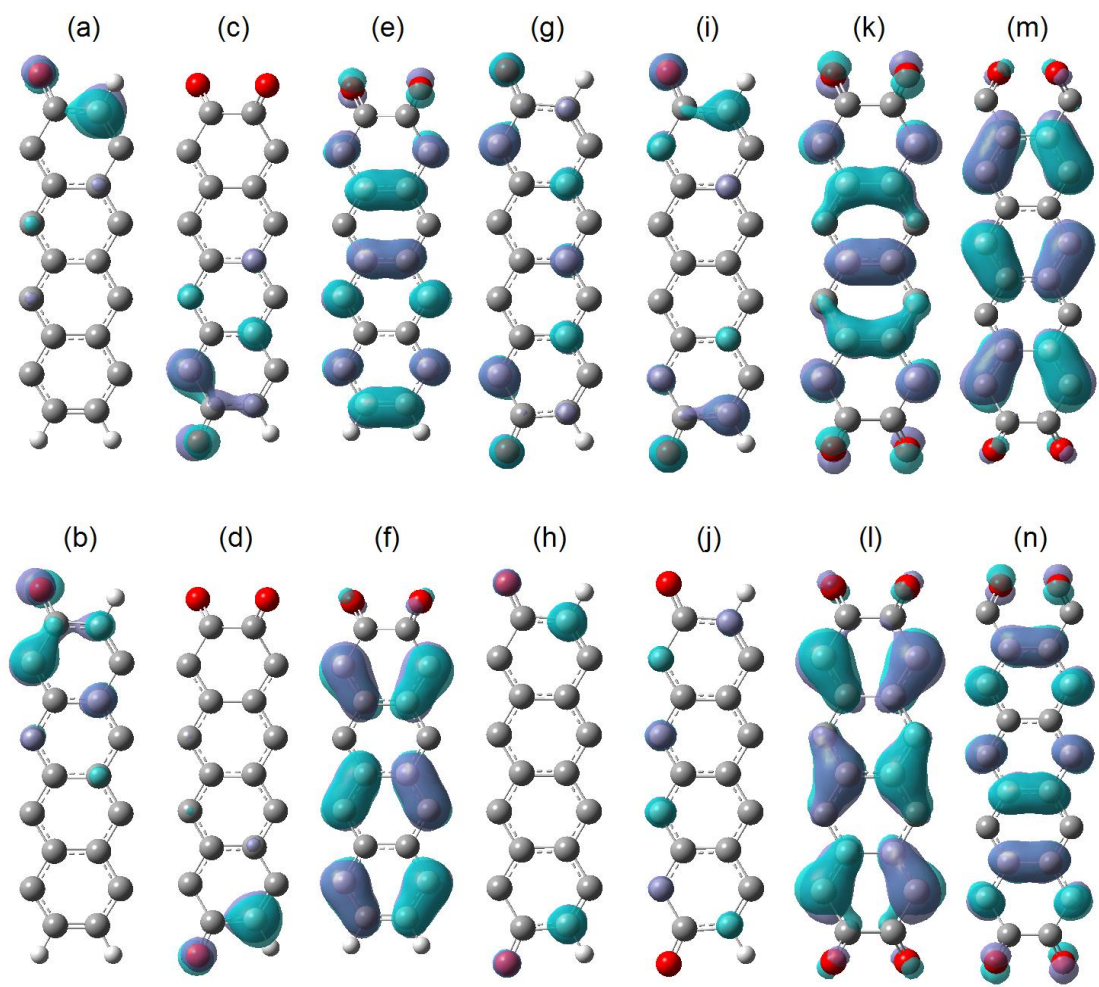

**Fig. 3.** (a) Real and (b) image part of frontier crystal orbitals of 1cell-1, (c) real and (d) image part of frontier crystal orbitals of 1cell-123, (e) HOCO and (f) LUCO of 1cell-13, (g) real and (h) image part of frontier crystal orbitals of 1cell-12 at α point, (i) real and (j) image part of frontier crystal orbitals of 1cell-12 at β point, (k) HOCO and (l) LUCO of 9-AGNR-O, (m) HOCO and (n) LUCO of 10-AGNR-O.

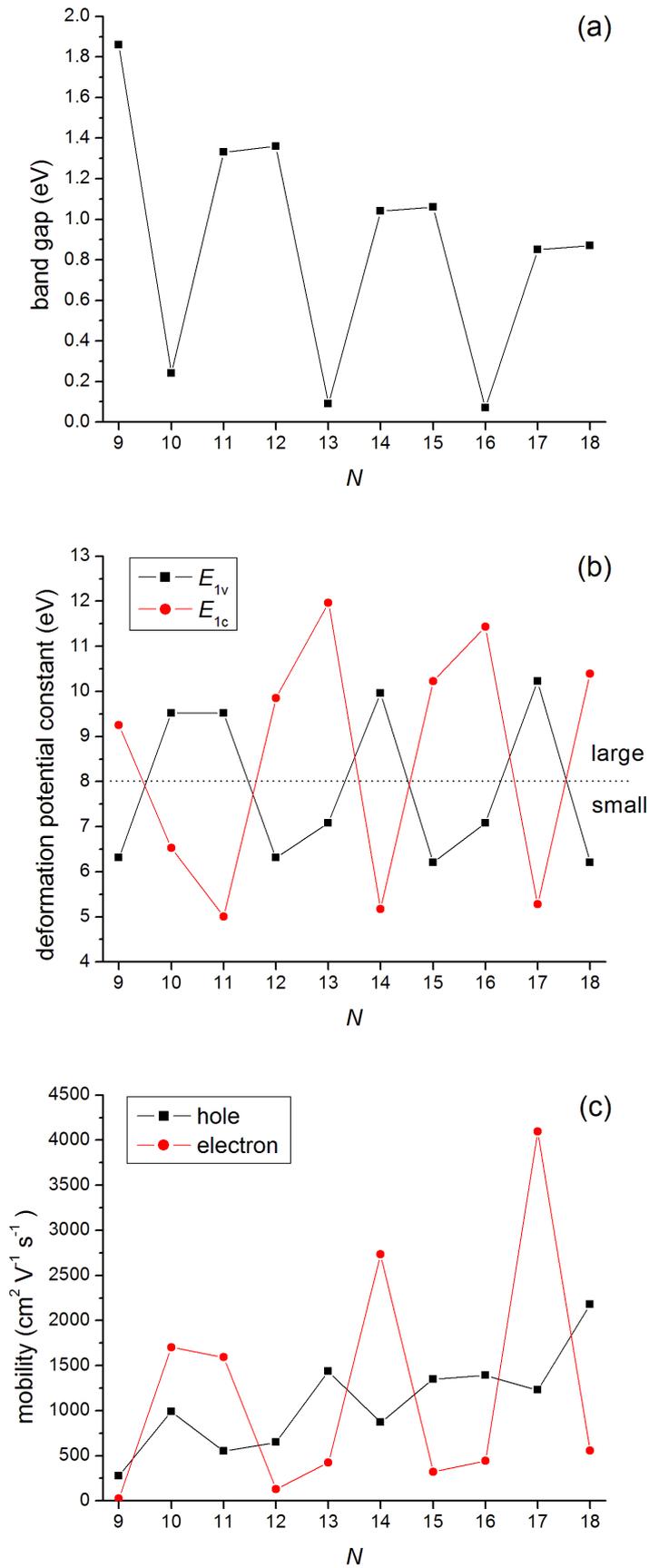

**Fig. 4.** (a) Band gaps, (b) deformation potential constants and (c) carrier mobilities of AGNR-O.